\documentstyle[12pt,epsfig]{article}
\def\Journal#1#2#3#4{{#1} {\bf #2}, #3 (#4)}

\def\PLB{{\em Phys. Lett.}  B}
\def\PRL{\em Phys. Rev. Lett.}
\def\PRD{{\em Phys. Rev.} D}
\def\NPA{{\em Nucl. Phys.} A}
\def\NPB{{\em Nucl. Phys.} B}
\def\PRC{{\em Phys. Rev.} C}

\begin{document}

\begin{titlepage}

\centerline{\large \bf $\Lambda(1405)$ and $\bar\Lambda(1405)$ in
$J/\psi$ four-body decays }

\vspace{1cm}

\centerline{Chiangbing Li and E. Oset}

\vspace{0.5cm}

\centerline{ Departamento de F\'{\i}sica Te\'{o}rica and IFIC, Centro Mixto Universidad de 
Valencia-CSIC,}
\centerline{ Institutos de Investigaci\'on de Paterna, Apdo. Correos 22085, 46071 Valencia, Spain}

\vspace{2.5cm}

\centerline{\bf Abstract}

\baselineskip 18pt

\vspace{0.5cm}

\noindent
We study the structure of the baryon resonances $\Lambda(1405)$ and $\bar\Lambda(1405)$ in 
$J/\psi$ four body decays $J/\psi\rightarrow \Sigma\bar \Sigma \pi\pi$ 
in the framework of a coupled channel chiral unitary approach. 
With still sufficient freedom for model parameters, the $\Lambda(1405)$ and $\bar\Lambda(1405)$
resonances are generated by simultaneously taking the meson baryon and meson anti-baryon 
final state interactions into account. The $\pi\Sigma$ ($\pi\bar\Sigma$) invariant mass distributions peak 
around 1410 MeV, which favors the assertion that the $\Lambda(1405)$ ($\bar\Lambda(1405)$)
is a superposition of the two $\Lambda(1405)$ ($\bar\Lambda(1405)$) states which dominantly 
couple to $\bar K N$ ($K\bar N$) and $\pi\Sigma$ ($\pi\bar\Sigma$),
respectively. We also calculate the amplitude for isospin $I=1$ 
which gives hints on a possible $I=1$ baryon resonance in the energy region of 
the $\Lambda(1405)$, which up to now has not been observed. 
\end{titlepage}

\baselineskip 18pt

{\flushleft \large \bf 1. Introduction }

\vspace{0.5cm}

The case of the $\Lambda(1405)$ is one of the examples of dynamically 
generated resonances which was already described within
scattering theory with coupled channels in \cite{dalitz}. More recently the advent of 
nonperturbative methods with input from chiral Lagrangian has set that
original idea on firmer grounds \cite{ksw,kww, om, phase, osetramos, magnetic, nieves,ref9}. 
The $\Lambda(1405)$ resonance, appearing about 30 MeV below the 
$\bar K N$ threshold plays a key role in the $\bar K N$ 
interaction and related processes and is a subject of
debate concerning its nature, whether
it is a genuine three quark system \cite{isgur, kimura} or a molecular-like meson-baryon 
bound state where chiral 
dynamics plays an important role. The recent discovery of the
pentaquark \cite{theta+} should stimulate again the debate on the nature of the $\Lambda(1405)$ since the
existence of that exotic state forces an interpretation of that baryon with 
at least five quarks \cite{5q1, 5q2},
although molecular structures with $K\pi N$ (heptaquark) have also been investigated \cite{7q1, 7q2}.
Within the chiral approach of \cite{ksw,kww, om, phase, osetramos, magnetic, nieves,ref9} 
the $\Lambda(1405)$ stands as a quasibound 
state of meson baryon, mostly $\bar K N$ and $\pi\Sigma$, which is also equivalent to 
five quark in the quark picture. The existence of the pentaquark makes 
more easily acceptable the idea of other {\it pentaquark} non-exotic state and vice versa.  
No doubt, explorations on the nature of the $\Lambda(1405)$ will provide
more clues to understand the non-perturbative nature of the QCD dynamics.

Chiral perturbation theory (ChPT) directly deals with hadron interactions in terms of meson-baryon
degrees of freedom.
As an effective field theory which incorporates the chiral symmetry of QCD, ChPT
has proved to be very successful in describing hadron interactions at low energies 
by expanding the chiral Lagrangian in powers of the hadron momentum. However, due to the problem 
of convergence of ChPT at relatively higher energies where most meson and baryon resonances appear,
the plain ChPT can do little for the description of resonances.
The lowest energy resonance in $\pi\pi$ scattering is the $\sigma$ which appears as a pole in the complex plane 
with a very large width. This is the case in the chiral unitary approach discussed below \cite{nucph, oop}
and is also the case in the $\pi\pi$ scattering amplitudes constructed by the Roy equation in \cite{gasser}, where the
width is as big as the real part of $\sim$500 MeV. Although this pole, far away from the real axis, has small repercussion 
at low energies where perturbative chiral calculations can be correctly applied, the existence of the 
complex variable theorem stating that a series expansion has a radii of convergence till 
the first singularity, already sets the 
limits on how far the perturbation expansion can be pushed.  

The chiral unitary coupled channels approach, which makes use of the 
standard ChPT Lagrangian together with an implicit or explicit expansion 
of $Re~T^{-1}$, instead of the $T$ matrix, 
has proved to be very successful in describing  meson meson \cite{nucph,oop} and meson 
baryon \cite{osetramos} interactions at higher energies. By employing the Chiral Lagrangian at the 
lowest order and solving the Bethe-Salpeter equation, this method was able to reproduce well the 
low-lying meson and baryon resonances in the PDG \cite{nucph,oop,excite}. 
When doing the extrapolation of ChPT at higher energies one usually neglects crossing symmetry since only the
right hand cut is used as a source of imaginary part of the amplitude and the left hand cut (unphysical cut)
is neglected. This, however, can be improved, as was done in \cite{nsd} for the meson meson interaction. Also
in \cite{om} a systematic method is proposed to also account for the left hand cut by including perturbative
crossed loop diagrams in the kernel of a dispersion relation for $T^{-1}$. In \cite{nsd} it was found that the
effect of the left hand cut was very small in a wide range of energies for the meson meson interaction below
$\sqrt{s}$=1.2 GeV. Similarly, in \cite{om} the effect of the left hand cut was estimated to be even 
smaller since crossed terms are reduced by factors of $(\frac{q}{2M})^2$ (with M the baryon mass) 
in the meson baryon interaction which are very small at the energies where the low lying baryon resonances
appear. The accuracy of the approximation neglecting the left hand cut has an important technical advantage since,
as proved in \cite{om}, the dispersion relation requires only the imaginary part of the meson baryon loop
(on shell part) and this leads to a Bethe-Salpeter (BS) type equation, identical to the one used in
\cite{osetramos, nucph}, where the kernel (potential) is needed only on shell. 
This converts the BS equation into an algebraic
equation, much as it happens with the use of a separable interaction, although there is
no need to define such a separable kernel.

With this chiral unitary approach the authors of \cite{nacher} calculated 
the photoproduction of the $\Lambda(1405)$ on the 
proton and nuclei and found different shapes of $\pi\Sigma$ invariant mass distributions
in different $\pi\Sigma$ charge channels, which was lately experimentally confirmed in \cite{lambex} and 
gave support to the assumption that the $\Lambda(1405)$ is a meson baryon loosely bound state.  
Additionally, it was found in \cite{2lamb} that the SU(3) symmetry breaking leads
to two poles of $\bar K N$ scattering matrix that might be responsible for the nominal 
$\Lambda(1405)$, one dominantly coupling to 
$\pi\Sigma$ and the other to $\bar K N$, and these poles 
are the mixing of the SU(3) singlet and octet. It was
concluded there that there are two $\Lambda(1405)$ resonances and the experimentally observed one is a 
superposition of the two states.    
However, whether the two poles really exist in the $\Lambda(1405)$ region is still unsolved experimentally. 
For this aim the authors of ref. \cite{klamb} suggested isolating the pole 
of $\Lambda(1405)$ that couples dominantly   
to $\bar K N$ in a photo-induced $K^*$ vector meson production process.
It will be interesting to further study the structure of 
$\Lambda(1405)$ in particular processes. 
  
In this work we propose to extract the structure of the $\Lambda(1405)$ in the 
decay processes $J/\psi\rightarrow \Sigma\bar\Sigma\pi\pi$ using the coupled channel chiral 
unitary approach to account for meson baryon and meson anti-baryon final state interactions (FSI). 
It is worth noting that in these processes the $\bar\Lambda(1405)$ could be generated 
through the $M\bar B$ FSI, together with the $\Lambda(1405)$ generated through the $MB$ FSI, 
and this would provide valuable information on the structure of the resonances. The $J/\psi$ four body decays
have been proposed to provide further information on the low-lying meson resonances in ref. \cite{lioset}, 
where the chiral unitary approach was employed to account 
for meson meson FSI without considering the $MB$ FSI. 
A natural continuation of the work of \cite{lioset} is to look for the meson baryon FSI that 
can lead to the formation of resonances, particularly those which in the chiral unitary approach
are dynamically generated. Experimental interest on the issue have been shown in 
a recent effort to search for the penta quark state 
in $J/\psi$ four body decays $J/\psi\rightarrow K_S^0 p K^- \bar n$ and 
$J/\psi\rightarrow K_S^0 \bar p K^+ n$ \cite{jpsipenta}.  

\vspace{0.8cm}

{\flushleft \large \bf 2. The model }

\vspace{0.5cm}

\begin{center}
\begin{figure}
\centerline{
\epsfig{file=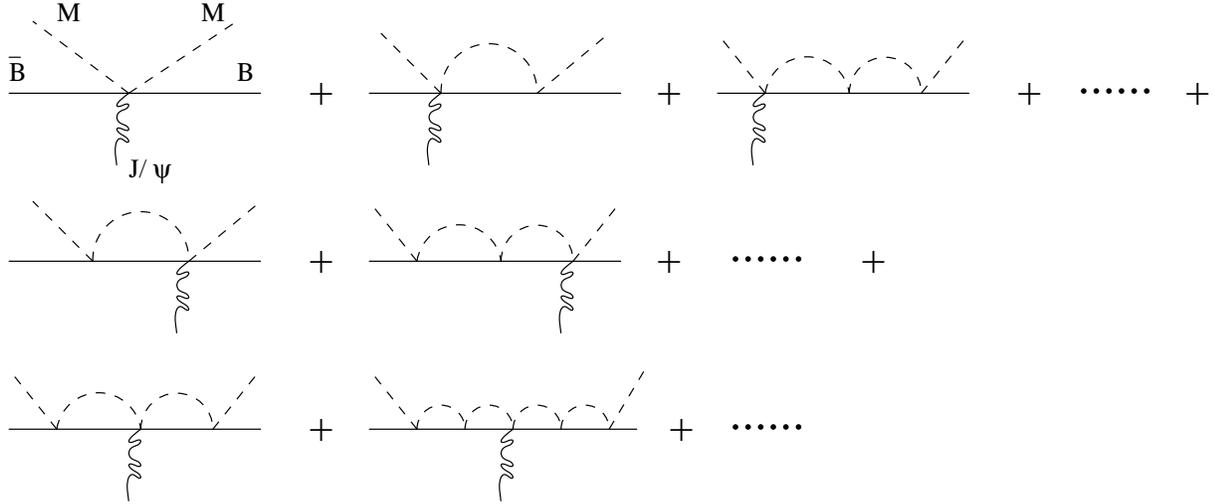,width=16.0cm,angle=0,clip=}}
\caption{\small Diagrams for $J/\psi\rightarrow B\bar B MM$ decays including the meson baryon and meson
anti-baryon final state interactions.  } 
\label{tgt}   
\end{figure}
\end{center}

We proceed now to construct the amplitudes for $J/\psi\rightarrow B \bar B MM$ simultaneously taking 
$MB$ and $M\bar B$ FSI into account, which is diagrammatically 
described in Fig. \ref{tgt}.

\vspace{-1.3cm}
Due to the lack of the
knowledge on the dynamics of charmonium decays, we employ the phenomenological 
Lagrangian used in \cite{lioset} which
incorporates SU(3) symmetry to account for the vertex of $J/\psi$ four body decays.  
Assuming that the $J/\psi$ is a SU(3) singlet \cite{book1}, the most general 
$B \bar B MM$ Lagrangian of SU(3) scalar nature without derivatives in the fields have the 
following possible structures:
\begin{eqnarray}
{\cal L}_1=g~Tr [\bar B\gamma^\mu B \Phi\Phi]\Psi_\mu, ~~~~~~~~
{\cal L}_2=g~Tr [\bar B\gamma^\mu\Phi B \Phi]\Psi_\mu, ~~~~\nonumber\\
{\cal L}_3=g~Tr [\bar B\gamma^\mu\Phi\Phi B]\Psi_\mu,  ~~~~~~~~
{\cal L}_4=g~Tr [\bar B\gamma^\mu B]Tr[\Phi\Phi]\Psi_\mu, 
\label{lags}
\end{eqnarray}
with $\Phi$, $B$ the ordinary SU(3) matrices for pseudoscalar mesons and $\frac{1}{2}^+$ baryons, 
respectively, $\Psi_\mu$ the $J/\Psi$ field and $g$ a constant to provide the right dimensions.
In constructing the effective Lagrangians we have imposed SU(3) symmetry to the Lagrangians together 
with the requirement of a minimum number of derivatives in the fields. We deliberately do not search for 
Lagrangians implementing chiral symmetry, which in view of the large number of particles involved and 
the fact that derivatives in the field are implied, would blow up the number of possible structures.  
Chiral Lagrangians are particularly useful to show how the interaction would change in 
the chiral limit when quark masses go to zero, but if the purpose is to have a parametrization of 
an amplitude accounting for the possible SU(3) structures, a procedure like the one done here
is sufficient within a limited range of energies. This is more the case in an approach like ours,
in which, as noted above, a factorization of on shell vertices is implicit in the loop, which will not
make derivative couplings to bring extra divergences.
Similar effective Lagrangians, without derivatives in the fields, have been used in related problems 
of $J/\psi$ decay, like $J/\psi\rightarrow \phi \pi\pi$ \cite{meissner, chiang}. 
As discussed in \cite{meissner}, the use of other Lagrangians involving derivative of 
the fields does not change the results and conclusions.  

We then take the Lagrangian of our problem as a linear combination of ${\cal L}_a$, $a=1,2,...,4$, 
\begin{eqnarray}
{\cal L}=\sum_{a=1}^4 x_a {\cal L}_a.
\label{lag}
\end{eqnarray}
This leads to the vertex for $J/\psi\rightarrow {(M \bar B)}_i (MB)_j$
\begin{eqnarray}
{\tilde V}_{ij}=-{\tilde c}_{ij} ~g~{\bar u}_j(p^\prime)\gamma^\mu v_i(p)\epsilon_\mu ({J/\psi}),
\label{gaugevij}
\end{eqnarray}
where we have already specified that we have a baryon anti-baryon production, rather than the baryon 
destruction and creation that one has for the meson
baryon amplitude. The eight coupled $(MB)_i$ (i=1, 2, ..., 8) channels that we consider 
are $K^- p$, ${\bar K}^0 n$, $\pi^0 \Lambda$, $\pi^0\Sigma^0$, $\eta\Lambda$, 
$\eta\Sigma^0$, $\pi^+\Sigma^-$ and $\pi^-\Sigma^+$. And the $(M \bar B)_i$ channels
are $K^+ \bar p$, $K^0 \bar n$, $\pi^0 \bar\Lambda$, $\pi^0\overline{\Sigma^0}$, $\eta\bar{\Lambda}$, 
$\eta\overline{\Sigma^0}$, $\pi^-\overline{\Sigma^-}$ and $\pi^+\overline{\Sigma^+}$. 
The $K\Xi$ state were shown in \cite{osetramos} to have no relevance in the $\Lambda(1405)$ dynamics. 
We list the ${\tilde c}_{ij} $ coefficients in Table I and we note that ${\tilde c}_{ji}={\tilde c}_{ij} $.

\vspace{0.5cm}

\begin{small}
\centerline{\small \bf Table I : ${\tilde c}_{ij} $ coefficients for $J/\psi\rightarrow {(M \bar B)}_i (MB)_j$ decays }
\begin{center}
\begin{tabular*}{15.8cm}{p{0.9cm}| p{1.3cm} p{1.3cm} p{1.5cm} p{1.5cm} p{1.5cm} p{1.5cm} p{1.3cm} p{1.3cm} 
  cccccccccc}
\hline 
\vspace{-0.4cm}\\
        & $~~~K^- p$ &  $~~~{\bar K}^0 n$ & $~~~\pi^0 \Lambda$ & $~~\pi^0\Sigma^0$ &$~~~~\eta\Lambda$ 
        & $~~~\eta\Sigma^0$ & $~\pi^+\Sigma^-$ & $~\pi^-\Sigma^+$   \\ \hline 
\vspace{0.05cm}	
$K^+ \bar p$ & $~~~~x_1+x_3+2x_4$ & $~~~~x_3$ &
$~~~\frac{x_1}{2\sqrt{3}}+\frac{x_2}{2\sqrt{3}}-\frac{x_3}{\sqrt{3}}$ 
        & $\frac{x_1}{2}+\frac{x_2}{2}$  & $~~-\frac{x_1}{6}+\frac{5x_2}{6}+\frac{x_3}{3}$  
        & $~-\frac{x_1}{2\sqrt{3}}+\frac{x_2}{2\sqrt{3}}$ & $~~~x_2$ & $~~~x_1$ \\ \hline  
\vspace{0.05cm}	
$K^0 \bar n$ & &$~~~x_1+x_3+2x_4$ & $~-\frac{x_1}{2\sqrt{3}}-\frac{x_2}{2\sqrt{3}}+\frac{x_3}{\sqrt{3}}$
                 &$\frac{x_1}{2}+\frac{x_2}{2}$ & $~~-\frac{x_1}{6}+\frac{5x_2}{6}+\frac{x_3}{3}$
                 &$~~\frac{x_1}{2\sqrt{3}}-\frac{x_2}{2\sqrt{3}}$ & $~~~x_1$ & $~~~x_2$  \\ \hline 
\vspace{0.05cm}		 
$\pi^0 \bar\Lambda$ &  &  & $\frac{x_1}{3}+\frac{x_2}{3}+\frac{x_3}{3}+2x_4$ & $~~~~~0$ & $~~~~~0$ 
                &$\frac{x_1}{3}+\frac{x_2}{3}+\frac{x_3}{3}$  & $~~~0$& $~~~0$\\ \hline 
\vspace{0.05cm}		
$\pi^0\overline{\Sigma^0}$ & & & & $x_1+x_2+x_3+2x_4$ 
                & $\frac{x_1}{3}+\frac{x_2}{3}+\frac{x_3}{3}$ & $~~~~~0$ & $~~~x_2$ & $~~~x_2$ \\ \hline 
\vspace{0.05cm}		
$\eta\bar\Lambda$   & & & & &  $x_1+x_2+x_3+2x_4$ 
                & $~~~~~0$ & $~~~\frac{x_1}{3}+\frac{x_2}{3}+\frac{x_3}{3}$ 
		& $~~~\frac{x_1}{3}+\frac{x_2}{3}+\frac{x_3}{3}$\\ \hline 
\vspace{0.05cm}
$\eta\overline{\Sigma^0}$  & & & & & & $\frac{x_1}{3}+\frac{x_2}{3}+\frac{x_3}{3}+2x_4$
                & $~~-\frac{x_1}{\sqrt{3}}+\frac{x_3}{\sqrt{3}}$ 
		& $\frac{x_1}{\sqrt{3}}-\frac{x_3}{\sqrt{3}}$  \\ \hline
\vspace{0.05cm}		 
$\pi^-\overline{\Sigma^-}$ & & & & & & & $~~~x_1+x_3+2x_4$ & $~~~x_2$   \\ \hline
\vspace{0.05cm}
$\pi^+\overline{\Sigma^+}$ & & & & & & & & $~~~x_1+x_3+2x_4$  \\ \hline 
\end{tabular*}
\label{coeff}
\end{center} 
\end{small}
In the rest frame of $J/\psi$, $\epsilon^0_r(J/\psi)$=0 for the three polarization vectors and 
eq.(\ref{gaugevij}) in the non-relativistic approximation for the nucleons can be written as 
\begin{eqnarray}
{\tilde V}_{ij}={\tilde c}_{ij} ~ g~ {\vec\sigma}\cdot{\vec\epsilon} (J/\psi)
\label{dvertex}
\end{eqnarray}
with $\vec\sigma$ the standard Pauli matrices.

We then construct the $J/\psi\rightarrow {(M \bar B)}_i (MB)_j$ amplitudes 
involving both $MB$ and $M\bar B$ FSI.
The decay amplitude with only $MB$ FSI can be written as  
\begin{eqnarray}
T_{ij}={\tilde V}_{ij}+\sum_k \, {\tilde V}_{ik}\,G_k\,t_{kj},
\label{tfirst}
\end{eqnarray}
where $t_{kj}$ are the scattering amplitudes for $(MB)_k\rightarrow (MB)_j$ 
which have been calculated in ref. \cite{osetramos}. The $(MB)_k$ loop integrals 
\begin{eqnarray}
G_{k} &=& i \, \int \frac{d^4 q}{(2 \pi)^4} \, \frac{M_k}{E_k
(\vec{q})} \,
\frac{1}{k^0 + p^0 - q^0 - E_k (\vec{q}) + i \epsilon} \,
\frac{1}{q^2 - m^2_k + i \epsilon} \nonumber \\
&=& \int \, \frac{d^3 q}{(2 \pi)^3} \, \frac{1}{2 \omega_k(q)}\,
\frac{M_k}{E_k (\vec{q})} \,
\frac{1}{p^0 + k^0 - \omega_k (\vec{q}) - E_k (\vec{q}) + i \epsilon}
\label{gprop}
\end{eqnarray}
only depend on the $MB$ invariant mass $p^0+k^0=\sqrt{s}$, where $p$ and $k$ are the momentum of the final 
baryon and meson, respectively, the masses of the particles and the cutoff of the meson three momentum
in the loop $q_{max}$. In Eq. (\ref{gprop}) $M_k, E_k$ are the mass and energy of the baryon in the loop,
respectively, and 
$\omega_k=\sqrt{{\vec q}^2+m_k^2}$ is the energy of the meson with the mass $m_k$ in the loop. 
The amplitudes $T_{ij}$ in Eq. (\ref{tfirst}) are functions of $\sqrt{s}$ once $q_{max}$ is 
fixed. In our calculation we take $q_{max}$=630 MeV. 
The value of $q_{max}$ was fixed in \cite{osetramos} in order to get an agreement of the theory with the lower 
energy parameters of the $K^- p$ interaction, and with this only parameter the 
$K^- p$ scattering cross sections and the invariant mass distribution of the $\Lambda (1405)$ 
resonance were well reproduced. Proper behavior of the loop functions requires that this cut off be reasonably 
bigger than the on shell momenta of the particles inside the loops. This puts some limit to
the range of energies where this can be used. A dimensional regularization of the loops was done 
in \cite{om} and \cite{excite} but, as shown in \cite{om}, the two procedures are practically equivalent
by establishing a correspondence between the cut off and the subtraction constant in dimensional regularization.
In \cite{excite} it was found that the cut off method could be safely used up to energies of 
$\sqrt{s}$=1670 MeV where the $\Lambda(1670)$ resonance is dynamically generated. 

The final amplitudes for $J/\psi$ four body
decays taking both $MB$ and $M\bar B$ FSI into account can be constructed as
\begin{eqnarray}
{\tilde T}_{ij}&=& T_{ij}+\, \sum_k \,{\bar t}_{ik}{\bar G}_k T_{kj}\nonumber \\
&=&{\tilde V}_{ij}+\, \sum_k \,{\tilde V}_{ik}G_k t_{kj}+\, \sum_k {\bar t}_{ik} {\bar G}_k {\tilde V}_{kj}
         +\sum_{kl} {\bar t}_{ik}{\bar G}_k {\tilde V}_{kl} G_l t_{lj},
\label{tfinal}
\end{eqnarray}
where ${\bar t}_{ik}$ is the scattering amplitudes for $(M\bar B)_i\rightarrow (M\bar B)_k$ and 
${\bar G}_k$ is the $(M\bar B)_k$ loop integration as given in Eq. (\ref{gprop}). 
It can be seen that Eq. (\ref{tfinal}) exactly corresponds to 
the diagrammatic description in Fig. \ref{tgt}.
Taking the same value as the $q_{max}$ of the $G$ loop integration for the 
cutoff in the $\bar G$ loop integration, 
the $\bar t$ and $\bar G$ matrices are identical to the $t$ and $G$ matrices derived
in ref. \cite{osetramos}, respectively, although they are functions of the $M\bar B$ 
invariant masses $\sqrt{s^\prime}$. Hence the amplitudes ${\tilde T}_{ij}$ 
in Eq. (\ref{tfinal}) are functions of $\sqrt{s}$ and $\sqrt{s^\prime}$.   

In Eqs. (\ref{tfirst}) and (\ref{tfinal}), 
the $MB$ and $M\bar B$ amplitudes in the loops are taken on shell. 
It was shown in \cite{osetramos} that the contribution of the off shell 
parts could be reabsorbed into a redefinition of coupling constants in $MB$ scattering.
Analogously, for the loop involving the vertex ${\tilde V}_{ij}$ in Eq. (\ref{tfirst}), which has 
a different structure from the one involving the $MB$ amplitude, 
the contribution of the off shell part in the loop 
could be absorbed by renormalizing the coupling constant $g$ in Eq. (\ref{lags}) since the integration for 
the off shell part in just one loop has the same structure as the tree diagram \cite{osetramos}. Similar 
arguments also apply for the off shell part of $M\bar B$ loops in Eq.(\ref{tfinal}).

In this paper we do not consider the MM interaction nor the $B\bar B$ interaction. 
The neglect of the $B\bar B$ interaction is justified since we are taking the phenomenological Lagrangians of
eq. (\ref{lags}) which were used in \cite{lioset} and fitted to the experimental data without taking into
account the $B\bar B$ interaction. Then the effect of $B\bar B$ interaction in the region of energies of
interest is accounted for phenomenologically in the couplings fitted to experiment in \cite{lioset}. This is,
however, not the case for the $MM$ interaction since this one was taken into account explicitly in
\cite{lioset}. However, the choice of different regions of phase space in \cite{lioset} and the present work
justifies the neglect of the $MB$ interaction in \cite{lioset} and of the $MM$ interaction in the present work.
The basic point is that the meson meson interaction is relatively weak except in the region of the resonances.
These would be in the case of meson meson interaction the $f_0(980)$ and $a_0(980)$ and to a much smaller
extend the broad $\sigma$, $\kappa$ of the $\pi\pi$, $\pi K$ interaction. In the case of $MB$ interaction, the
resonance of relevance in the region studied here is the $\Lambda(1405)$. 
When we concentrate in a narrow region around 1405 MeV for the invariant  mass of $\pi\Sigma$ and in
addition in the same region of energies for the invariant mass of the $\pi\bar\Sigma$, one selects a very
narrow region of the four body phase space where there is a large enhancement because of the double resonance
structure of $\pi\Sigma$ and $\pi\bar\Sigma$. However, this region of phase space contains the whole range of
invariant masses of the meson meson combinations, and then, the possible effects of the $MM$ resonances is
diluted since the $MM$ resonance region will only appear in a very narrow region of the phase space where one
is integrating.
A practical manifestation of this disentangling of the interactions when one look at peaks of resonances in
particular channels is seen in \cite{nacher} where one studies the $\gamma p\rightarrow K^+ \pi\Sigma$ with
$\pi\Sigma$ in the $\Lambda(1405)$ resonance region. The $MB$ interaction is considered there but the $MM$
interaction is neglected and the predictions show good agreement with experimental results \cite{lambex}.
Conversely, in \cite{marco} the same reaction was used, paying attention to the meson meson interaction alone,
in order to evaluate cross sections for the production of scalar mesons.

\vspace{0.8cm}

{\flushleft \large \bf 3. Results and discussions }

\vspace{0.5cm}

The mass distribution  of the decays $J/\psi\rightarrow {(M \bar B)}_i (MB)_j$
with respect to the $MB$ and $M\bar B$ invariant masses, which is particularly suited to search for
resonances, can be written as \cite{phasespace}
\begin{eqnarray}
\frac{d^2 \Gamma_{ij}}{dM_{I} dM^\prime_{I}}&=&\frac{1}{(2\pi)^8 
2M_J}\frac{\pi^3}{2M_J^2 M_{I}^2 {M^\prime _{I}}^2} ~4M_{I} M^\prime _{I} M
M^\prime ~~\lambda^{\frac{1}{2}}({M_I}^2,m^2,M^2)\nonumber \\ 
&&\lambda^{\frac{1}{2}}({M_I^\prime}^2,{m^\prime}^2,{M^\prime}^2)
\lambda^{\frac{1}{2}}(M_{I}^2,{M^\prime_{I}}^2,M_{J}^2)
~~\overline{\sum} \sum~|{\tilde T}_{ij}(M_{I},M^\prime_{I})|^2,
\label{distrib}
\end{eqnarray}
with $M_{I}$ and $M^\prime_{I}$ being the $MB$ and $M\bar B$ invariant masses, respectively,
$m$ and $m^\prime$ the meson masses in final states, $M$ and $M^\prime$ the $B$ and $\bar B$ masses, 
respectively, $M_J$ the mass of $J/\psi$,    
and $\lambda(x^2,y^2,z^2)$ the Kaellen function. Eq. (\ref{distrib}) differs slightly from
eq. (46) of \cite{phasespace} because of our different normalization of the fields and the
$\tilde T$ matrix. It should be stated that the simple form of this equation
holds because of our neglect of the meson meson and baryon antibaryon
interactions. We shall come back to this point at the end of the results
section.

We then perform calculations for the decays $J/\psi\rightarrow \Sigma\bar\Sigma \pi\pi$ to search for 
the $\Lambda(1405)$ and ${\bar\Lambda}(1405)$.
We have the $x_i$ with $i=1, 2, ..., 4$ in Eq. (\ref{lag}) as the model parameters.
Similarly to what was done in \cite{lioset}, we define the ratios $r_i=\frac{x_i}{x_3}$ with $i$=1, 2, 4,
which were evaluated by fitting the experimental data of $J/\psi\rightarrow p\bar p \pi^+ \pi^-$ decay. 
It was shown in \cite{lioset} that the parameter $r_1$ influences the  
shape of the $\pi\pi$ spectrum 
of $J/\psi\rightarrow p\bar p \pi^+\pi^-$ at higher energies, but its contribution could be 
included in the variation of $r_4$. The parameter 
$r_2$ does not influence the $J/\psi\rightarrow p\bar p \pi^+\pi^-$ but 
plays an important role for the decays considered here. 
In our calculations $r_4$ is given the values of 
ref. \cite{lioset}, which reproduces the empirical $\pi\pi$ spectrum in 
$J/\psi\rightarrow p\bar p \pi^+\pi^-$ decay. We take $r_4$=0.2, which has been used in ref. \cite{lioset}. 
As for $r_2$ we take it as a free parameter and vary it in a wide range from $0.1\sim 2.0$. Although the
branching ratio obtained varies much, the important thing from where the conclusions will be drawn is
the shape of the distribution and this does not depend on the precise value of $r_2$.
Similarly, changes in $r_2$ for the value $r_4$=-0.27, which was also able to well reproduce the data for 
$J/\psi\rightarrow p\bar p \pi^+\pi^-$, as is shown in ref. \cite{lioset} 
within a reasonable range, also does not change the qualitative character of the results 
for the channels considered here. In the following we take $r_4$=0.2 and $r_2$=0.6
to give characteristic descriptions for the generation of $\Lambda(1405)$ and ${\bar\Lambda}(1405)$ 
in $J/\psi\rightarrow \Sigma\bar\Sigma \pi\pi$ decays. Then we have the value of the constant
$g_\alpha=(x_3+2x_4)g=1.1\times 10^{-6}~ MeV^{-2}$, which determines the shape of the $\pi\pi$ spectrum
and the width of the decay $J/\psi\rightarrow p\bar p \pi^+\pi^-$ \cite{lioset}. 

\begin{center}
\begin{figure}
\centerline{
\epsfig{file=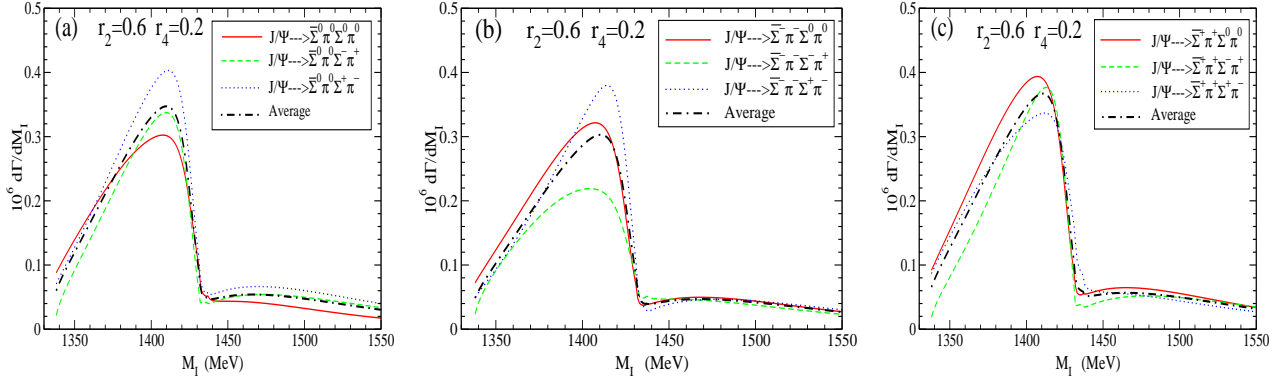,width=18.0cm,angle=0,clip=}}
\caption{\small (Color online) The $\pi\Sigma$ invariant 
mass distributions to account for $\Lambda(1405)$ for 
$J/\psi\rightarrow \Sigma {\bar \Sigma} \pi \pi$ 
decays with the model parameters $r_2=0.6$, $r_4 = 0.2$, where the involved meson and baryons 
are assigned physical
masses.} 
\label{plotall}   
\end{figure}
\end{center}

We present the $\pi\Sigma$ invariant mass distributions for the nine 
$J/\psi\rightarrow \Sigma\bar\Sigma \pi\pi$ decay channels in Fig. \ref{plotall}, where baryons 
and mesons are assigned physical masses. It can be seen that the shapes of the distributions are 
quite different from each other, where all coupled channels collaborate to build up the $\Lambda(1405)$
resonance. The figure has labels like $J/\psi\rightarrow \overline{\Sigma^-}\pi^-\Sigma^0\pi^0$ 
etc, which means that one measures simultaneously
the $\overline{\Sigma^-}\pi^-$ and $\Sigma^0\pi^0$ and integrates eq. (\ref{distrib}) over the 
invariant mass of $\overline{\Sigma^-}\pi^-$ to provide the $\Sigma^0\pi^0$ mass distribution shown in 
the figure. Similarly we could plot the figure for $\frac{d\Gamma}{dM_I^\prime}$ for the invariant mass of the  
$\overline{\Sigma^-}\pi^-$  and other channels integrating over the invariant mass of the $\pi\Sigma$ 
system. Certainly channels like $J/\psi\rightarrow \overline{\Sigma^-}\pi^-\Sigma^-\pi^+$ would 
have identical distributions for the $\overline{\Sigma^-}\pi^-$ or $\Sigma^-\pi^+$ invariant masses.
For other combinations one finds small differences between $\bar \Sigma\pi$ and $\Sigma\pi$
mass distributions because of small mass differences of the involved $\Sigma$ baryons and $\pi$ mesons.   

The results in Fig. \ref{plotall} can be better understood with the isospin decomposition for $\pi\Sigma$
\begin{eqnarray}
&&|\pi^0\Sigma^0\rangle = \sqrt{\frac{2}{3}}~|2,0\rangle ~-~ \sqrt{\frac{1}{3}}~|0,0\rangle~,\nonumber \\
&&|\pi^+\Sigma^-\rangle = -~\sqrt{\frac{1}{6}}~|2,0\rangle ~-~\sqrt{\frac{1}{2}}~|1,0\rangle 
                          ~-~ \sqrt{\frac{1}{3}}~|0,0\rangle~,\nonumber \\
&&|\pi^-\Sigma^+\rangle = ~-~\sqrt{\frac{1}{6}}~|2,0\rangle ~+~\sqrt{\frac{1}{2}}~|1,0\rangle 
                          ~-~ \sqrt{\frac{1}{3}}~|0,0\rangle		
\end{eqnarray} 
and for $\pi\bar\Sigma$
\begin{eqnarray}
&&|\pi^0\overline{\Sigma^0}\rangle = \sqrt{\frac{2}{3}}~{|2,0\rangle} ~-~
\sqrt{\frac{1}{3}}~{|0,0\rangle}~,\nonumber \\
&&|\pi^-\overline{\Sigma^-}\rangle = -~\sqrt{\frac{1}{6}}~|2,0\rangle ~+~\sqrt{\frac{1}{2}}~|1,0\rangle 
                          ~-~ \sqrt{\frac{1}{3}}~|0,0\rangle~,\nonumber \\
&&|\pi^+\overline{\Sigma^+}\rangle = -~\sqrt{\frac{1}{6}}~|2,0\rangle -\sqrt{\frac{1}{2}}~|1,0\rangle 
                          ~-~ \sqrt{\frac{1}{3}}~|0,0\rangle		
\label{cg}
\end{eqnarray}
with $|\pi^+\rangle =-|1,1\rangle$, $|\Sigma^+\rangle =-|1,1\rangle$ and 
$|\overline{\Sigma^-}\rangle =-|1,1\rangle$, we have the amplitudes for the particular 
$J/\psi\rightarrow (M\bar B)_i (MB)_j$ decays
\begin{eqnarray}
&&{\tilde T}_{44} ~=~ \frac{2}{3}~T^{(2)}~+~\frac{1}{3}~T^{(0)}~,\nonumber\\
&&{\tilde T}_{47} ~=~-~ \frac{1}{3}~T^{(2)}~+~\frac{1}{3}~T^{(0)}~,\nonumber\\
&&{\tilde T}_{48} ~=~-~ \frac{1}{3}~T^{(2)}~+~\frac{1}{3}~T^{(0)}~,\nonumber\\
&&{\tilde T}_{77} ~=~ \frac{1}{6}~T^{(2)}~-~\frac{1}{2}~T^{(1)}~+~\frac{1}{3}~T^{(0)}~,\nonumber\\
&&{\tilde T}_{78} ~=~ \frac{1}{6}~T^{(2)}~+~\frac{1}{2}~T^{(1)}~+~\frac{1}{3}~T^{(0)}~,\nonumber\\
&&{\tilde T}_{88} ~=~ \frac{1}{6}~T^{(2)}~-~\frac{1}{2}~T^{(1)}~+~\frac{1}{3}~T^{(0)}~, 
\label{tiso}
\end{eqnarray}
and we note ${\tilde T}_{ij}={\tilde T}_{ji}$. 
It can be seen that the shapes for the particular decay channels show evidence of 
some isospin breaking which appears naturally in our framework because of the different 
masses of the members of the same isospin multiplet. This is the case, for instance, of the channels
$J/\psi\rightarrow \overline{\Sigma^0}\pi^0\Sigma^-\pi^+$ and 
$J/\psi\rightarrow \overline{\Sigma^0}\pi^0\Sigma^+\pi^-$ (channels 47 and 48) 
in Fig. \ref{plotall}(a), which according to eq. (\ref{tiso}) should give the same distributions.
Similarly, the distributions of the channels $J/\psi\rightarrow \overline{\Sigma^-}\pi^-\Sigma^0\pi^0$
in Fig. \ref{plotall}(b) and $J/\psi\rightarrow \overline{\Sigma^+}\pi^+\Sigma^0\pi^0$ in Fig.
\ref{plotall}(c) (channels 74 and 84, respectively) should also be equal.
The larger differences that one observes in Fig. \ref{plotall}(b) between 
$J/\psi\rightarrow \overline{\Sigma^-}\pi^-\Sigma^-\pi^+$
and $J/\psi\rightarrow \overline{\Sigma^-}\pi^-\Sigma^+\pi^-$ (channels 77 and 78, respectively),
are due to mixed terms of the type $Re(T^{(1)}{T^{(2)}}^*)$ and $Re(T^{(1)}{T^{(0)}}^*)$, implying,
in this case, that these terms are larger than the differences due to isospin breaking.

In Fig. \ref{plotall} we have also calculated the averaged cross sections. By using again eqs.
(\ref{tiso}) it is easy to see that all the mixed terms $Re(T^{(i)}{T^{(j)}}^*)$ in the modulus squared of
the amplitudes cancel in these averages and one has only contributions of ${|T^{(i)}|}^2$.  
These averages should not be the same since they come from different combination of 
${|T^{(i)}|}^2$, but the fact that they are not very different indicates that they are all dominated by the 
${|T^{(0)}|}^2$ component, which appear in all of them with the same weight, $\frac{1}{3}{|T^{(0)}|}^2$, 
and that the other isospin components are much smaller. Hence, this averaged distributions is 
the closest thing one can get experimentally for the shape of the $\Lambda(1405)$ resonance. 
>From the position of the peak of the distributions
around 1410 MeV and the width of around 60 MeV, the results imply that we have  
a superposition of the two $\Lambda(1405)$ ($\bar\Lambda(1405)$) resonances 
(with poles at (1390-i60) MeV and (1426-i16) MeV) found in \cite{2lamb}. 

\begin{center}
\begin{figure}
\centerline{
\epsfig{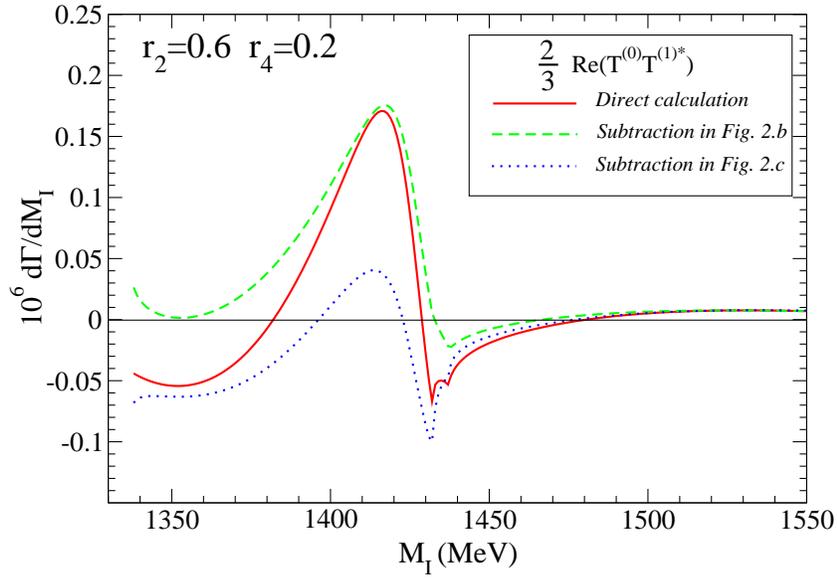}}
\caption{\small (Color online) The $\pi\Sigma$ invariant mass distribution for 
the $\frac{2}{3}Re(T^{(0)}{T^{(1)}}^*)$ of 
$J/\psi\rightarrow \Sigma {\bar \Sigma} \pi \pi$. The solid line denote the results
through direct theoretical calculations and the dashed and dotted lines are the results 
calculated by subtracting the $\pi\Sigma$ invariant mass distributions of the particular decays
$J/\psi\rightarrow \overline {\Sigma^-} \pi^- \Sigma^-\pi^+$ and
$J/\psi\rightarrow \overline {\Sigma^-} \pi^- \Sigma^+\pi^-$ in 
Fig. \ref{plotall}(b) and $J/\psi\rightarrow \overline {\Sigma^+} \pi^+ \Sigma^-\pi^+$ and
$J/\psi\rightarrow \overline {\Sigma^+} \pi^+ \Sigma^+\pi^-$ in Fig. \ref{plotall}(c), respectively. } 
\label{t0t1}   
\end{figure}
\end{center}

\begin{center}
\begin{figure}
\centerline{
\epsfig{file=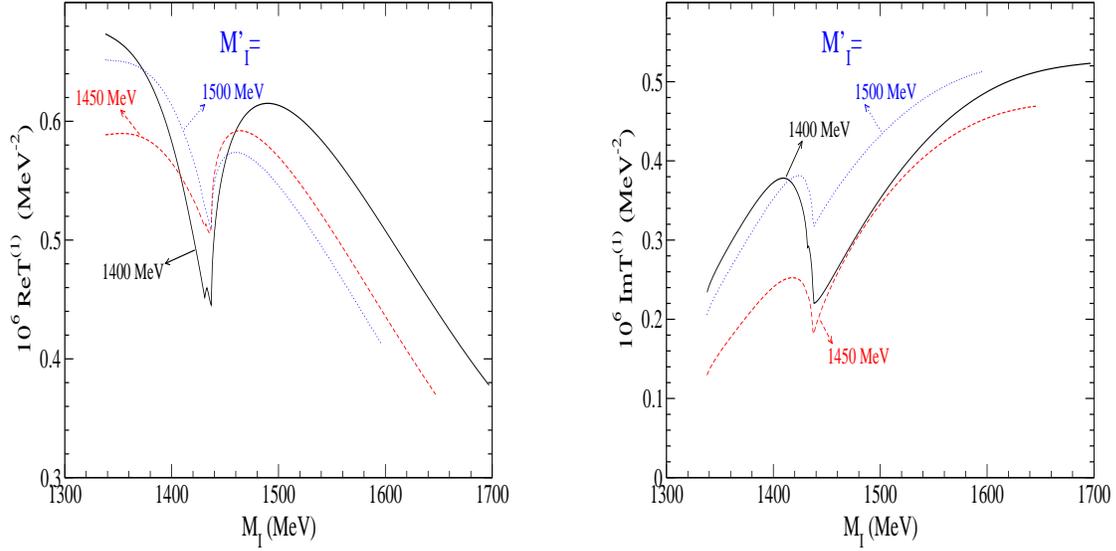,width=16.0cm,angle=0,clip=}}
\caption{\small (Color online) The real part and imaginary part of the 
$T^{(1)}$ amplitude as a function of $M_I$ for
different values of the $\pi\bar\Sigma$ invariant masses.} 
\label{t1}   
\end{figure}
\end{center}

Now we turn to another interesting potential use of these reactions. It was found 
in \cite{om} that there was a pole of $I=1$ close to the $\bar K N$ threshold which would correspond to
a new resonance not accounted for in the Particle Data Book. Under certain circumstances, with smaller
degree of SU(3) breaking, it was also found in \cite{2lamb} using the approach of \cite{excite}. 
It is interesting to see what these reactions can say to this respect. For instance, 
from Eq. (\ref{tiso}) we see 
\begin{eqnarray}
|{\tilde T}_{78}|^2~-~|{\tilde T}_{77}|^2 ~=~
|{\tilde T}_{87}|^2~-~|{\tilde T}_{88}|^2 ~=~\frac{2}{3}~Re(T^{(0)}{T^{(1)}}^*) 
+\frac{1}{3}~Re(T^{(1)}{T^{(2)}}^*),  
\label{subt}
\end{eqnarray}
where the $T^{(1)}{T^{(2)}}^*$ term is negligible.
This means that the subtraction of the mass distributions of the charged decay channels
$J/\psi\rightarrow \overline{\Sigma^-}\pi^-\Sigma^+\pi^-$ and 
$J/\psi\rightarrow \overline{\Sigma^-}\pi^-\Sigma^-\pi^+$ can give hints on the $T^{(1)}$
amplitude, given that the amplitude $T^{(0)}$ can in principle be derived by averaging the mass 
distributions for particular $J/\psi\rightarrow \bar\Sigma\pi\Sigma\pi$ decays. 
In Fig. \ref{t0t1} we plot two different results. One is done by calculating $T^{(0)}$ and $T^{(1)}$ from 
combination of the amplitudes given in eqs. (\ref{tiso}). Then a mass distribution is generated 
by replacing $\overline{\sum} \sum~|{\tilde T}|^2~$ in eq. (\ref{distrib}) with
$\frac{2}{3}Re(T^{(0)}{T^{(1)}}^*)$. The second calculation corresponds to what an experimentalist 
could do by subtracting the two mass distributions corresponding to the 
$J/\psi\rightarrow \overline{\Sigma^-}\pi^-\Sigma^+\pi^-$ and 
$J/\psi\rightarrow \overline{\Sigma^-}\pi^-\Sigma^-\pi^+$ channels in Fig. \ref{plotall}(b) or
$J/\psi\rightarrow \overline{\Sigma^+}\pi^+\Sigma^+\pi^-$ and 
$J/\psi\rightarrow \overline{\Sigma^+}\pi^+\Sigma^-\pi^+$ channels in Fig. \ref{plotall}(c).
The considerable different magnitude between the results of the two subtractions in Fig. \ref{t0t1}
is the manifestation of the isospin symmetry breaking that we have in this approach. 
With isospin symmetry the two subtractions give approximately the same results. The $T^{(1)}$ amplitude
could in principle be extracted from Fig. \ref{t0t1} and 
be compared to the theoretically calculated one. For this purpose, in Fig. \ref{t1}  
we show the real part and imaginary part of $T^{(1)}$ directly calculated in our model
for some fixed values of the $M\bar B$ invariant masses. 
It can be seen that the shape of the $T^{(1)}$ does not 
qualitatively change with respect to the selected values of $M_I^\prime$ and there exists nontrivial
cusps around the energy $M_I$=1420 MeV, which may give some hints for the possible 
approximate resonant structure predicted in \cite{om, 2lamb}. Calling the attention that
the reactions discussed here bear potentially valuable information of this $I=1$ amplitude is one of the
purposes of the present work. 

The calculations have been done using the cut off of $q_{max}$=630 MeV used in \cite{osetramos}. 
This cut off was the only free parameter there to fit different $\bar K N$ cross sections, threshold values 
plus the $\Lambda(1405)$ shape. The freedom in this parameter is very small if a good fit to these data is
demanded. Changes in the parameter from 630 MeV to 620 MeV or 640 MeV is as much as one can afford. We have  
reevaluated our results with these new values of the parameter and find changes of about 10\% in the cross
sections. These should be considered as theoretical uncertainties from this source.  

\begin{center}
\begin{figure}
\centerline{
\epsfig{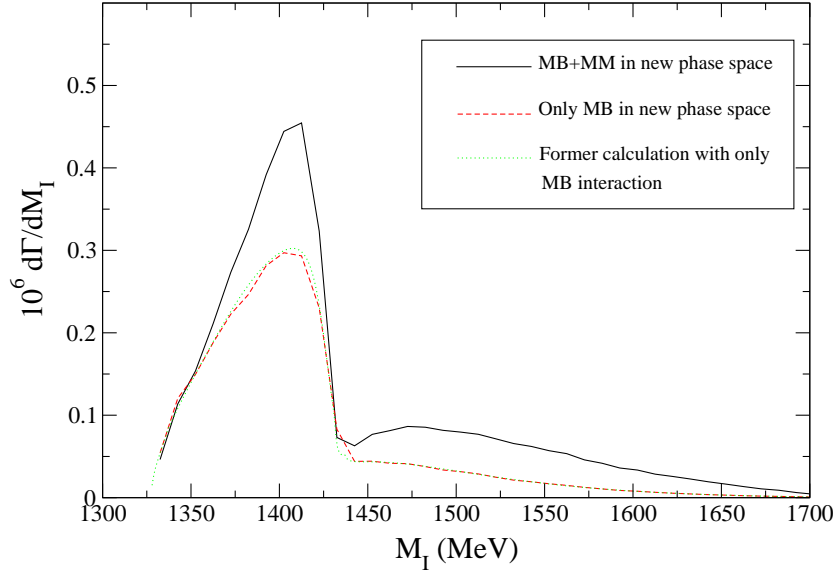}}
\caption{\small (Color online) The $\pi\Sigma$ invariant mass distribution of the 
$J/\psi\rightarrow \overline{\Sigma^0}\pi^0\Sigma^0\pi^0$ channel calculated in new 
phase space of eq. \ref{newphase} with 
Monte Carlo method. Solid line: the result in new phase space considering both meson baryon and
meson meson interactions; Dashed line: the result in new phase space considering only meson
baryon interactions; Dotted line: the result calculated with eq. (\ref{distrib}) considering only 
meson baryon interaction.} 
\label{nim44all}   
\end{figure}
\end{center}

So far we have taken into account the meson baryon (meson anti-baryon) interaction only. 
Since in \cite{lioset},
by studying the same problem, we took into account the meson meson interaction, with some additional 
work we can consider the two sources of interactions to see how our results can 
change with the inclusion of the 
meson meson interaction. The new calculation has been done by adding to the diagrams of Fig. \ref{tgt} the 
rescattering diagrams of Fig. 1 of ref. \cite{lioset}. This means we would consider the diagrams with meson 
rescattering stemming from the first diagram in Fig. \ref{tgt} of the present paper. The neglect of the 
rescattering terms from the other diagrams can be justified in the fact that 
the two pions can be produced at reasonably 
large distances where this interaction should be very weak. We must now modify 
eq. (\ref{distrib}) since the 
$\tilde T$ matrix now depends on other variables. The standard formula in this case is
\begin{eqnarray}
d\Gamma=\frac{(2\pi)^4}{2M_J}~|\tilde T|^2 ~d\Phi(P; p_1, p_2, k_1, k_2)
\label{newphase}
\end{eqnarray}
with the four-body phase space being
\begin{eqnarray}
d\phi(P; p_1, p_2, k_1, k_2)&=&\delta^4(P-p_1-p_2-k_1-k_2)~\frac{d^3p_1}{(2\pi)^3}\frac{M_1}{E(p_1)}~\nonumber\\
&&\frac{d^3p_2}{(2\pi)^3}\frac{M_2}{E(p_2)}~\frac{d^3k_1}{(2\pi)^3}\frac{1}{2\omega(k_1)}~
\frac{d^3k_2}{(2\pi)^3}\frac{1}{2\omega(k_2)},
\end{eqnarray}
where $P$ is the four-momentum of $J/\psi$, $p_i$ and $k_i$ (i=1, 2) the four-momenta
of the involved baryons and mesons, respectively, $M_i$ the baryon masses and $E(p_i)=\sqrt{M_i^2+p_i^2}$
and $\omega(k_i)=\sqrt{m_i^2+k_i^2}$, with $m_i$ being meson 
masses, the baryon and meson energies, respectively.  
We perform the integrations with the Monte Carlo method. We show the results in 
Fig. \ref{nim44all} for the channel 
$J/\psi\rightarrow\overline{\Sigma^0}\pi^0\Sigma^0\pi^0$ (results are similar for other channels). 
We observe that the strength in the region of energies above 1405 MeV gets considerably increased, smearing 
the meson resonance contribution in a large phase space region. We also observe that the shape of the 
$\Lambda (1405)$ peak
is not changed but there is an extra strength which accounts for about 30\% of the total. 
This reflects a constructive 
interference between the resonance amplitude and the background from the meson meson interaction.
This extra contribution coming from the consideration of the meson meson interaction 
changes the strength of the peak in about the same amount in the different channels 
and does not affect the qualitative nature of the conclusions drawn here.

\vspace{0.8cm}

{\flushleft \large \bf 3. Summary}

\vspace{0.5cm}

In summary, we investigate the structure of the baryon resonances $\Lambda(1405)$ and $\bar\Lambda(1405)$
in $J/\psi$ four body decays $J/\psi\rightarrow \Sigma\bar\Sigma\pi\pi$. It is shown that 
$\Lambda(1405)$ and $\bar\Lambda(1405)$ are generated by simultaneously taking 
the FSI of $\pi\Sigma$ and $\pi\bar\Sigma$ into account, which is calculated in the framework 
of the chiral unitary approach. By averaging the three $\pi\Sigma$ ($\pi\bar\Sigma$) 
mass distributions in either of the three 
plots in Fig. \ref{plotall} we get the real shape for the nominal 
$\Lambda(1405)$ ($\bar\Lambda(1405)$) resonance,
which peaks around 1410 MeV and is a superposition of the two $\Lambda(1405)$ states discussed in
\cite{2lamb}. On the other hand, the subtraction of the charged decay channels 
in either Fig. \ref{plotall}(b) or Fig. \ref{plotall}(c) gives hints on the possible $I=1$ resonance in the energy region
of $\Lambda(1405)$, which up to now has not been observed. From an
experimental point of view, although all the nine particular 
$J/\psi\rightarrow \Sigma\bar\Sigma\pi\pi$ decays were considered here for a theoretical analysis,  
the three of them in either Fig. \ref{plotall}(b) or Fig. \ref{plotall}(c) are practically adequate to extract the 
structure of $\Lambda(1405)$, $\bar\Lambda(1405)$ and the possible $I=1$ resonance in this region.
It is worth noting that our theoretical calculations were done with still sufficient freedom of 
model parameters. However, the fact that the variation of the parameters within a reasonable range does not change 
the qualitative feature of the results, together with the success of the chiral unitary 
approach in past work, sets the predictions made here on firmer grounds.  
Experimental data on these channels would be 
most welcome to further
fix the model parameters and to get refined predictions. No doubt, the experimental investigations 
on the proposed $J/\psi$ four body decays
will provide interesting information of the structure of $\Lambda(1405)$ and $\bar\Lambda(1405)$ resonances
and valuable test for the approaches employed here.     

\vspace{3cm}

{\flushleft{\bf Acknowledgments}}

We would like to thank M. J. Vicente Vacas for valuable discussions.
C. L. acknowledges the hospitality of the University of Valencia where this work has been done and 
financial support from the Ministerio de Educacion in the program 
Doctores y Tecn\'{o}logos extranjeros under contract number SB2000-0233.
This work is partly supported by
DGICYT project M2003-00856 and the EU network EURIDICE contract HPRN-CT-2002-00311. 

\end{document}